\documentclass [12pt,a4paper]{article}
\usepackage{amssymb}
\renewcommand{\>}{\rangle}

\def\bbbr{{\mathbb R}}
\def\eps{\varepsilon}
\def\pont{\,\cdot\,}
\def\Tr{\mbox{Tr}\,}
\def\Diag{\mbox{Diag}\,}
\def\tr{\mbox{Tr}\,}
\def\im{\mbox{i}}
\def\Scal{\mbox{Scal}\,}
\def\aa{\alpha}

\def\J{{\mathbb J}}

\def\bL{{\mathbb L}}
\def\bR{{ \mathbb R}}

\def\fel{\textstyle{\frac{1}{2}}}
\def\iM{{\cal M}}
\def\ffi{\varphi}
\newtheorem{thm}{Theorem}[section]
\begin{document}
\ \vskip 1cm
\centerline{\LARGE Covariance and Fisher information}
\medskip
\centerline{\LARGE in quantum mechanics}
\bigskip
\centerline{\large D\'enes Petz}
\bigskip
\centerline{Department for Mathematical Analysis}
\centerline{Budapest University of Technology and Economics}
\centerline{ H-1521 Budapest XI., Hungary}
\bigskip\bigskip
\noindent

\textbf{Abstract}
Variance and Fisher information are ingredients of the Cram\'er-Rao
inequality. We regard Fisher information as a Riemannian metric on
a quantum statistical manifold and choose monotonicity under coarse 
graining as the fundamental property of variance and Fisher
information. In this approach we show that there is a kind of dual
one-to-one correspondence between the candidates of the two
concepts. We emphasis that Fisher informations are obtained from
relative entropies as contrast functions on the state space and argue 
that the scalar curvature might be interpreted as an uncertainty density 
on a statistical manifold.

On the one hand standard quantum mechanics is a statistical theory, on
the other hand, there is a so-called geometrical approach to
mathematical statistics \cite{Am, Ce}. In this paper the two topics
are combined and the concept of covariance and Fisher information 
is studied from an abstract poit of view. We start with the 
Cram\'er-Rao inequality to realize that the two concepts are very
strongly related. What they have in common is a kind of monotonicity
property under coarse grainings. (Formally the monotonicity of
covariance is a bit difference from that of Fisher information.)
Monotone quantities of Fisher information type determine a 
superoperator $\J$ which gives immediately a kind of generalized
covariance. In this way a one-to-one correspondence is established 
between the candidates of the two concepts.
In the paper we prove a Cram\'er-Rao type inequality in the setting
of generalized variance and Fisher information. Moreover, we argue
that the scalar curvature of the Fisher information Riemannian metric
has a statistical interpretation. This gives interpretation of an
earlier formulated but still open conjecture on the monotonicity
of the scalar curvature. 

\section{The Cram\'er-Rao inequality for an introduction}

The Cram\'er-Rao inequality belongs to the basics of estimation theory
in mathematical statistics. Its quantum analog was discovered 
immediately after the foundation of mathematical quantum estimation 
theory in the 1960's, see the book \cite{He} of Helstrom, or the
book \cite{Ho} of Holevo for a rigorous summary of the subject. Although 
both the
classical Cram\'er-Rao inequality and its quantum analog are as trivial
as the Schwarz inequality, the subject takes a lot of attention
because it is located on the highly exciting boundary of statistics,
information and quantum theory.

As a starting point we give a very general form of the quantum
Cram\'er-Rao inequality in the simple setting of finite dimensional
quantum mechanics. For $\theta\in (-\eps, \eps)\subset \bbbr$ a statistical
operator $D_\theta$ is given and the aim is to estimate the value 
of the parameter $\theta$ close to $0$. Formally $D_\theta$ is an $n
\times n$ positive semidefinite matrix of trace 1 which describes a
mixed state of a quantum mechanical system and we assume that $D_\theta$
is smooth (in $\theta$). In our approach we deal with mixed states 
contrary to several 
other authors, see \cite{Fu}, for example. Assume that an estimation 
is performed by the measurement of a selfadjoint matrix $A$ playing the 
role of an observable. $A$ is called {\bf locally unbiased estimator} if
\begin{equation}\label{E:lue}
\frac{\partial}{\partial \theta}\Tr D_\theta A\Big|_{\theta=0}=1\,.
\end{equation} 
This condition holds if $A$ is an {\bf unbiased estimator} for $\theta$,
that is
\begin{equation}
\Tr D_\theta A =\theta \qquad (\theta \in (-\eps,\eps)).
\end{equation}
To require this equality for all values of the parameter is a serious
restriction on the observable $A$ and we prefer to use the weaker
condition (\ref{E:lue}).

Let $\ffi_0[\pont,\pont]$ be an inner product on the linear space of
selfadjoint matrices. $\ffi_0[\pont,\pont]$ depends on the density 
matrix $D_0$, the notation reflects this fact. When $D_\theta$ is smooth 
in $\theta$, as already was assumed above, the correspondence
\begin{equation}\label{E:func}
B \mapsto \frac{\partial}{\partial \theta}\Tr D_\theta B\Big|_{\theta=0}
\end{equation}
is a linear functional on the selfadjoint matrices and it is of the
form $\ffi_0[B,L]$ with some $L=L^*$. From (\ref{E:lue}) and (\ref{E:func})
we have $\ffi_0[A,L]=1$ and the Schwarz inequality yields
\begin{equation}\label{E:CR}
\ffi_0[A,A] \ge \frac{1}{\ffi_0[L,L]}\,.
\end{equation} 
This is the celebrated inequality of Cram\'er-Rao type for the locally
unbiased estimator. We want to interprete the left-hand-side as a 
{\bf generalized variance} of $A$.
The right-hand-side of (\ref{E:CR}) is independent of the estimator
and provides a lower bound for the generalized variance. The 
denominator $\ffi_0[L,L]$ appears to be in the role of Fisher information 
here. We call it {\bf quantum Fisher information} with respect to the
generalized variance $\ffi_0[\pont,\pont]$. This quantity depends on the
tangent of the curve $D_\theta$.

We want to conclude from the above argument that whatever Fisher 
information and generalized variance are in the quantum mechanical
setting, they are very strongly related. In an earlier work 
(\cite{PD2, PD3}) we used a monononicity condition to make a limitation
on the class of Riemannian metrics on the state space of a quantum
system. The monotone metrics are called Fisher information quantities
in this paper. Now we observe that a similar monotonicity property
can be used to get a class of bilinear forms, we call the elements of
this class generalized variances. The usual variance of two observables
is included but many other quantities as well. We descibe a one-to-one
correspondence beween variances and Fisher informations. The correspondence
is given by a superoperator $\J$ which appears immediately in the analysis
of the inequality (\ref{E:CR}).

Since the sufficient and necessary condition for the equality in the
Schwarz inequality is well-known, we are able to analyze the case of 
equality in (\ref{E:CR}). The condition for equality is
$$
A=\lambda L
$$
for some constant $\lambda \in \bbbr$. On the $n \times n$ selfadjoint
matrices we have two inner products: $\ffi_0[\pont,\pont]$ and $\<A,B\>:=
\Tr AB$. There exists a linear operator $\J_0$ on the selfadjoint 
matrices such that
$$
\ffi_0[A,B] =\Tr A\J_0 (B)\,.
$$
Therefore the necessary and sufficient condition for equality in 
(\ref{E:CR}) is
\begin{equation}\label{E:EQ}
\dot{D}_0:=\frac{\partial}{\partial \theta} D_\theta 
\Big|_{\theta=0}=\lambda^{-1} \J_0(A)\,.
\end{equation}
Therefore there exists a unique locally unbiased estimator $A=\lambda\J_0^{-1}
(\dot{D}_0)$, where the number $\lambda$ is chosen such a way that the
condition (\ref{E:lue}) should be satisfied.

\section{Coarse graining and Fisher information}

In the simple setting in which the state is described by a density
matrix, a coarse graining is an affine mapping sending density matrices
into density matrices. Such a mapping extends to all matrices and
provides a positivity and trace preserving linear transformation.
A common example of coarse graining sends the density matrix $D_{12}$
of a composite system $1+2$ into the (reduced) density matrix $D_1$ of 
component 1. (There are several reasons to assume completely positivity
about a coarse graining but now we do not consider this issue.)

Assume that $D_\theta$ is a smooth curve of density matrices with
tangent $A:=\dot{D}_0$ at $D_0$. The quantum Fisher information 
$F_D(A)$ is an information quantity associated with the pair $(D_0, A)$,
it appeared in the Cram\'er-Rao inequality above and the Fisher
information gives a bound for the (generalized) variance of a locally
unbiased estimator. Let now $\alpha$ be a coarse graining. Then $\alpha 
(D_\theta)$ is another curve in the state space. Due to the linearity of 
$\alpha$, the tangent at $\alpha(D_0)$ is $\alpha(A)$. As it is usual in
statistics, information cannot be gained by coarse graining, therefore 
we expect that the Fisher information at the density matrix $D_0$ in 
the direction $A$ must be larger than the Fisher information at
$\alpha(D_0)$ in the direction $\alpha(A)$. This is the {\bf monotonicity
property of the Fisher information} under coarse graining:
\begin{equation}\label{E:FM}
F_D(A) \ge F_{\alpha(D)}(\alpha(A))
\end{equation}
Although we do not want to have a concrete formula for the quantum 
Fisher information, we require that this monotonicity condition must hold.
Another requirment is that $F_D(A)$ should be quadratic in $A$, in
other words there exists a nondegenerate real bilinear form 
$\gamma_D(A,B)$ on the selfadjoint matrices such that
\begin{equation}\label{E:BFM}
F_D(A)=\gamma_D(A,A).
\end{equation}
The requirements (\ref{E:FM}) and (\ref{E:BFM}) are strong enough
to obtain a reasonable but still wide class of possible quantum Fisher
informations. 

We may assume that 
\begin{equation}\label{E:Jdef}
\gamma_D(A,B)=\Tr A\J_D^{-1}(B^*).
\end{equation}
for an operator $\J_D$ acting on matrices. (This formula
expresses the inner product $\gamma_D$ by means of the
Hilbert-Schmidt inner product and the positive linear 
operator $\J_D$.) In terms of the operator $\J_D$ the 
monotonicity condition reads as
\begin{equation}\label{E:Fmon}
\alpha^* \J_{\alpha(D)}^{-1}\alpha  \le \J_D^{-1}
\end{equation}
for every coarse graining $\alpha$. ($\alpha^*$ stand for the adjoint
of $\alpha$ with respect to the Hilbert-Schmidt product. Recall that
$\alpha$ is completely positive and trace preserving if and only if
$\alpha^*$ is completely positive and unital.) On the other hand the latter
condition is equivalent to
\begin{equation}\label{E:fimon1}
\alpha \J_{D}\alpha^*  \le \J_{\alpha(D)}\, .
\end{equation}

We proved the following theorem in \cite{PD2}, see also
\cite{PD4}.

\begin{thm}\label{T:mon}
If for every density matrix $D$ a positive definite bilinear form 
$\gamma_D$ is given such that (\ref{E:FM}) holds for all completely
positive coarse grainings $\alpha$ and $\gamma_D(A,A)$ is continuous
in $D$ for every fixed $A$, then there exists a unique operator
monotone function $f: \bbbr^+\to \bbbr$ such that $f(t)=tf(t^{-1})$
and $\gamma_D(A,A)$ is given by the following prescription.
$$
\gamma_D(A,A)=\Tr A\J_D^{-1}(A)\qquad\mbox{and}\qquad
\J_D=\bR_D^{1/2}f(\bL_D\bR_D^{-1})\bR_D^{1/2}\,,
$$
where the linear transformations $\bL_D$ and  $\bR_D$ acting on matrices
are the left and right multiplications, that is
$$
\bL_D(X)=DX \qquad\mbox{and}\qquad \bR_D(X)=XD\,.
$$
\end{thm}

Although the statement of the theorem seems to be rather complicated,
the formula for  $F_D(A)=\gamma_D(A,A)$ becomes simpler when $D$ and 
$A$ commute. On the subspace $\{A:  AD=DA\}$ the left multiplication
$\bL_D$ coincides with the right one $\bR_D$ and $f(\bL_D\bR_D^{-1})=
f(1)$. Therefore we have
\begin{equation}\label{E:kom}
F_D(A)=\frac{1}{f(1)}\Tr D^{-1}A^2\qquad \mbox{if}\qquad AD=DA.
\end{equation}
Under the hypothesis of commutation the quantum Fisher information is
unique up to a constant factor. (This fact reminds us the Cencov
uniqueness theorem in the Kolmogorovian probability, \cite{Ce}.
According to this theorem the metric on finite probability spaces is
unique when monotonicity under Markovian kernels is posed.)
We say that the quantum Fisher information is {\bf classically
Fisher-adjusted} if 
\begin{equation}\label{E:FA}
F_D(A)=\Tr D^{-1}A^2\qquad \mbox{when}\qquad AD=DA.
\end{equation}
This means that we impose the normalization $f(1)=1$ on the operator
monotone function. In the sequel we always assume this condition.

Via the operator $\J_D$, each monotone Fisher information determines
a quantity
\begin{equation}\label{E:fi}
\varphi_D[A,A]:= \Tr A\J_D(A) 
\end{equation}
which could be called {\bf generalized variance}. According to 
(\ref{E:fimon1}) this possesses the monotonicity property
\begin{equation}\label{E:fimon2}
\varphi_D[\alpha^*(A),\alpha^*(A)] \le \varphi_{\alpha(D)}[A, A]\, .  
\end{equation}
Since (\ref{E:Fmon}) and (\ref{E:fimon1}) are equivalent we observe a 
{\bf one-to-one correspondence between monotone Fisher informations and 
monotone generalized variances}. Any such variance has the property
$\varphi_{D}[A, A]=\Tr DA^2$ for commuting $D$ and $A$. The examples
below show that it is not so generally.

The analysis in \cite{PD2} led to the fact that among all monotone
quantum Fisher informations there is a smallest one which corresponds
to the function $f_m(t)=(1+t)/2$. In this case 
\begin{equation}\label{E:minF}
F_D^{\min}(A)=\Tr A L=\Tr DL^2,\qquad \mbox{where}\qquad DL+LD=2A.
\end{equation}
For the purpose of a quantum Cram\'er-Rao inequality the minimal quantity
seems to be the best, since the inverse gives the largest lower bound.
In fact, the matrix $L$ has been used for a long time under the name
of {\bf symmetric logarithmic derivative}, see \cite{Ho} and \cite{He}.
In this example the generalized covariance is
\begin{equation}\label{E:minFv}
\ffi_D[A,B] =\fel \Tr D(AB+BA)
\end{equation}
and we have
\begin{equation}\label{E:minFj}
\J_D(A)=\fel(DA+AD)\qquad \mbox{and} \qquad
\J_D^{-1}(A)=L=2\int_0^\infty e^{-tD}Ae^{-tD}\,dt
\end{equation}
for the superoperator $\J$ of the previous section.

The set of invertible $n \times n$ density matrices is a manifold of
dimension $n^2-1$. Indeed, parametrizing these matrices by $n-1$ real
diagonal entries and $(n-1)n/2$ upper diagonal complex entries we have
$n^2-1$ real parameters which run over an open subset of the Euclidean
space $\bbbr^{n^2-1}$. Since operator monotone function are smooth
(even analytic), all the quantities $\gamma_D$ in Theorem \ref{T:mon}
endow the manifold of density matrices with a Riemannian structure.

\section{Garden of monotone metrics}

All the monotone quantum Fisher information quantities in the range of
the previous theorem are depending smoothly on the footpoint density
$D$ and hence they endow the state space with a {\bf Riemannian structure}.
In particular, the Riemannian geometry of the minimal Fisher information 
was the subject of the paper \cite{Di}.
 
It is instructive to consider the state space of a 2-level quantum system in 
details. Dealing with $2\times 2$ density matrices, we conveniently use
the so-called Stokes parametrization.
\begin{equation} \label{eq:Stokes}
D_x = \fel (I + x_1\sigma_1+ x_2\sigma_2+ x_3 \sigma_3)\equiv
\fel (I+x\cdot \sigma)
\end{equation}
where $\sigma_1, \sigma_2, \sigma_3$ are the Pauli matrices and
$(x_1, x_2, x_3) \in \bbbr^3$ with $x^2_1 + x^2_2 + x^2_3 \le 1$.
A monotone Fisher information on $\iM_2$ is rotation invariant in the
sense that it depends only on $r=\sqrt{x^2+y^2+z^2}$ and splits into
radial and tangential components as follows.
\begin{equation} \label{eq:3.7}
ds^2={1 \over 1-r^2}dr^2+{1 \over 1+r}g\Big({1-r \over 1+r}\Big)dn^2\,,
\quad {\rm where}\quad g(t)={1 \over f(t)}\,.
\end{equation}

The radial component is independent of the function $f$. (This fact is
again a reminder of the Cencov uniqueness theorem.) 
The  limit of the tangential component exists in (\ref{eq:3.7}) 
when $r \to 1$ provided that $f(0)\ne 0$. In this way the standard
Fubini-Study metric is obtained on the set of pure states, up to a
constant factor. (In case of larger density matrices, pure states form a 
small part of the topological boundary of the invertible
density matrices. Hence, in order to speak about the extension of a 
Riemannian metric on invertible densities to pure states, a rigorous 
meaning of the extension should be given. This is the subject of the paper
\cite{Sudar-1996}, see also \cite{PD4}.) Besides minimality the radial
extension yields another characterization of the minimal quantum Fisher 
information, see \cite{PD4}.

\begin{thm}
Among the monotone quantum Fisher informations the minimal one (given
by (\ref{E:minF})) is characterized by the properties that it is
classically Fisher-adjusted (in the sense of (\ref{E:FA})) and its
radial limit is the Fubini-Study metric on pure states.
\end{thm}

We note that in the minimal case $f_m(t)=(t+1)/2$ we have constant
tangential component in (\ref{eq:3.7}):
\begin{equation} \label{eq:3.8}
ds^2={1 \over 1-r^2}dr^2+dn^2 \,.
\end{equation}

The metric (\ref{E:minF}) is widely accepted in the role of quantum Fisher 
information, see \cite{oebn}. However, some other operator monotone 
functions may have importance. Let us see first the other extreme.
According to \cite{PD2} there is a largest metric among all monotone
quantum Fisher informations and this corresponds to the function $f_M(t)
=2t/(1+t)$. In this case 
\begin{equation}\label{E:maxF}
\J_D^{-1}(A)=\fel(D^{-1}A+AD^{-1})\qquad \mbox{and} \qquad
F_D^{\max}(A)=\Tr D^{-1}A^2.
\end{equation}
The maximal metric cannot be extended to pure states.

It can be proved that the function
\begin{equation} \label{E:efek}
f_\beta(t)=\beta(1-\beta)\frac{(x-1)^2}{(x^\beta-1)(x^{1-\beta}-1)}
\end{equation}
is operator monotone. This was done for the case $0 < \beta < 1$ in
\cite{peha} and the case $-1 < \beta < 0$ was treated in \cite{hape}.
(The operator monotonicity follows also from (\ref{E:Ruskfor}) below.)
We denote by $F^\beta$ the corresponding Fisher information metric.
When $A=\im [D,B]$ is orthogonal to the commutator of the footpoint
$D$ in the tangent space, we have
\begin{equation} \label{E:WYD}
F^\beta_D(A)=\frac{1}{2 \beta(1-\beta)}\tr\big([D^\beta,B][D^{1-\beta}, 
B]\big).
\end{equation}
Apart from a constant factor this expression is the skew information
proposed by Wigner and Yanase some time ago (\cite{WYD}). In
the limiting cases $\beta \to 0$ or $1$ we have
$$
f_0(x)=\frac{1-x}{\log x}
$$ 
and the corresponding metric
\begin{equation} \label{E:KM}
K_D(A,B):=  \int_0^\infty \tr A (D+t)^{-1}B(D+t)^{-1}\,dt
\end{equation}
is named after Kubo, Mori, Bogoliubov etc. The Kubo-Mori inner product
plays a role in quantum statistical mechanics (see \cite{FS}, for
example). In this case
\begin{equation} \label{E:KMJ}
\J^{-1}(B)=  \int_0^\infty (D+t)^{-1}B(D+t)^{-1}\,dt\quad\hbox{and}\quad
\J(A)=  \int_0^1 D^{t}A D^{1-t}\,dt\,.
\end{equation}
Therefore the corresponding generalized variance is
\begin{equation} \label{E:KMvar}
\varphi_D(A,B)=  \int_0^1 \tr A D^{t}BD^{1-t}\,dt\,.
\end{equation}
 
Beyond the affine parametrization of the set of density matrices, the 
exponential parametrization is another possibility: Any density matrix
is written in a unique way in the form $e^H/\tr e^H$, where $H$ is a 
selfadjoint traceless matrix. In the affine parametrization the
integral (\ref{E:KM}) gives the metric and (\ref{E:KMvar}) is the 
corresponding variance. If we change for the exponential parametrization,
the role of the two formulas is interchanged: integral (\ref{E:KM}) gives 
the variance and (\ref{E:KMvar}) is the metric. (The reason for this fact
that the change of the coordinates is described by $\J$ from (\ref{E:KMJ}).)
The affine and exponential parametrization is the subject of the paper
\cite{hase} and the characterization of the Kubo-Mori metric in
\cite{GS} is probably another form of the duality observed between
(\ref{E:KM}) and (\ref{E:KMvar}).

\section{The Cram\'er-Rao inequalities revisited}

Let $\iM:=\{D_\theta: \theta \in G\}$ be a smooth $m$-dimensional 
manifold, parametrized in such a way that $0\in G\subset \bbbr^m$.
A (locally) unbiased estimator of $\theta$ at $\theta=0$ is a collection 
$A=(A_1,\dots,A_m)$ of selfadjoint matrices, such that
\begin{enumerate}
\item[(i)] $ \Tr D_0A_i=0$ for all $1\le i \le m$,
\item[(ii)] $\frac {\partial}{\partial \theta_i}\Tr D_\theta A_j|_{\theta_i=0}= 
\delta_{ij}$ for all $ i,j=1,\dots,m$.
\end{enumerate}

Suppose a generalized variance $\varphi_0$ is given. Then the generalized 
covariance matrix of the estimator $A$ is a positive definite matrix, 
defined by $\varphi_0[A]_{ij}=\varphi_0[A_i,A_j]$. If
$$
\frac {\partial}{\partial \theta_i}\Tr D_\theta B\Big|_{\theta_i=0}=
\varphi_0[L_i,B]
$$
determines the logarithmic derivatives $L_i$, then
$$
\varphi_0[A_i,L_j]=\delta_{ij}\qquad (i,j=1,\dots,m).
$$
This orthogonality relation implies a matrix inequality for the Gram matrices
which is an inequality of  Cram\'er-Rao type. 

\begin{thm}\label{cr-rao} Let $A=(A_1,\dots,A_m)$ be a locally (at $\theta=0$) 
unbiased estimator of $\theta$, moreover $L_i$ and $\varphi_0$ be as
above. Then
$$
\varphi_D[A]\geq \left(\big(\varphi_0[L_i,L_j]\big)_{ij}\right)^{-1}
$$
in the sense of the order on positive definite matrices. 
\end{thm}

The proof is rather simple if we use the block matrix method. Let $X$
and $B$ be $m \times m$ matrices with $n\times n$ entries and assume
that all entries of $B$ are constant multiples of the unit matrix. 
($A_i$ and $L_i)$ are $n \times n$ matrices.) If
$\alpha$ is a completely positive mapping on $n\times n$ matrices, then
$\tilde\alpha:=\Diag(\alpha,\dots, \alpha)$ is a positive mapping on
block matrices and $\tilde\alpha (BX)=B\tilde\alpha(X)$. This implies that
$\tr X\alpha(X^*)B \ge 0$ when $B$ is positive. Therefore the $m\times
m$ ordinary matrix $M$ which has $ij$ entry
$$
\tr (X\tilde\alpha(X^*))_{ij}
$$
is positive. In the sequel we restrict ourselves for $m=2$ for the sake
of simplicity and apply the above fact to the case
$$
X=\left[
\begin{array}{cccc}
A_1&0&0&0\\
A_2&0&0&0\\
L_1&0&0&0\\
L_2&0&0&0\\
\end{array}
\right] \quad \hbox{and}\quad \alpha=\J_D\,.
$$
Then we have
$$
M=\left[\begin{array}{cccc} \tr A_1\J_D(A_1)&\tr A_1\J_D(A_2)&
\tr A_1\J_D(L_1)&\tr A_1\J_D(L_2)\\
\tr A_2\J_D(A_1)&\tr A_2\J_D(A_2)&
\tr A_2\J_D(L_1)&\tr A_2\J_D(L_2)\\ 
\tr L_1\J_D(A_1)&\tr L_1\J_D(A_2)&
\tr L_1\J_D(L_1)&\tr L_1\J_D(L_2)\\
\tr L_2\J_D(A_1)&\tr L_2\J_D(A_2)&
\tr L_2\J_D(L_1)&\tr L_2\J_D(L_2)\\
\end{array}
\right]\ge 0
$$

Now we rewrite the matrix $M$  in terms of a generalized variance
$\varphi_0$ and apply the orthogonality assumption. We get
$$
M=\left[\begin{array}{cccc} \varphi_0[ A_1, A_1]&\varphi_0[A_1,A_2]&
1 & 0\\
\varphi_0[A_2, A_1]& \varphi_0[A_2, A_2]&
0 & 1\\ 
1 & 0 &
\varphi_0[L_1,L_1]& \varphi_0[L_1, L_2]\\
0& 1&
\varphi_0[L_2, L_1]& \varphi_0[L_2,L_2]\\
\end{array}
\right]\ge 0
$$
Since the positivity of a block matrix
$$
M=\left[\begin{array}{cc}M_1&I\\I&M_2\\\end{array}\right]=
\left[\begin{array}{cc}\varphi_D[A]&I\\I&\big(\varphi_D[L_i,L_j]_{ij})
\\ \end{array}\right]
$$
implies $M_1 \ge M_2^{-1}$ we have exactly the statement of our
Cram\'er-Rao inequality.

\section{Statistical distinguishability and uncertainty}

Assume that a manifold $\iM:=\{D_\theta: \theta \in G\}$ of density 
matrices is given together a statistically relevant Riemannian metric 
$\gamma_d$. We do not give a formal definition of such a metric. What we 
have in mind is the property that given two points on the manifold 
their geodesic distance is interpreted as the statistical distinguishability
of the two density matrices in some statistical procedure.

Let $D_0 \in \iM$ be a point on our statistical manifold. The geodesic ball
$$
B_\eps (D_0):=\{D \in \iM: d(D_0,D) < \eps\}
$$
contains all density matrices which can be distinguished by an effort smaller
than $\eps$ from the fixed density $D_0$. The size of the inference region
$B_\eps (D_0)$  measures the statistical uncertainty at the density $D_0$. 
Following {\bf Jeffrey's rule} the size is the volume measure determined by the
statistical (or information) metric. More precisely, it is better
to consider the asymptotics of the volume of $B_\eps (D_0)$ as $\eps \to 0$. 
According to differential geometry
\begin{equation}
Vol \big(B_\eps (D_0)\big)=C_n \eps^n -\frac{C_n}{6(n+2)}\Scal (D_0)\eps^{n+2}+
o(\eps^{n+2}),
\end{equation}
where $n$ is the dimension of our manifold, $C_n$ is a constant (equals to the
volume of the unit ball in the Euclidean $n$-space) and $Scal$ means the scalar
curvature, see 3.98 Theorem in \cite{GHL}. In this way, the scalar curvature 
of a statistically relevant Riemannian metric might be interpreted as the 
{\bf average statistical uncertainty} of the density matrix (in the given 
statistical manifold). This interpretation becomes particularly interesting
for the full state space endowed by the Kubo-Mori inner product as a 
statistically relevant Riemannian metric.

Let $\iM$ be the manifold of all invertible $n \times n$ density matrices. The
Kubo-Mori (or Bogoliubov) inner product is given by
\begin{equation}
\gamma_D(A,B)=\Tr (\partial_A D)( \partial_B \log D).
\end{equation}
In particular, in the affine parametrization we have
\begin{equation}
\gamma_D(A,B)=\int_0^\infty \Tr A(D+t)^{-1}B(D+t)^{-1},
\end{equation}
see \cite{PD1}. On the basis of numerical evidences it was {\bf conjectured}
in \cite{PD1} that the scalar curvature which is a statistical uncertainity
is monotone in the following sense. For any coarse graining $\alpha$
the scalar curvature at a density $D$ is smaller than at $\alpha(D)$.
The average statistical uncertainty is increasing under coarse graining.
Up to now this conjecture has not been proven mathematically. Another form
of the conjecture is the statement that along a curve of Gibbs states
$$
\frac{ e^{-\beta H}}{\Tr e^{-\beta H}}
$$
the scalar curvature changes monotonly with the inverse temperature 
$\beta \ge 0$, that is, {\bf the scalar curvature is monotone decreasing 
function of $\beta$}.  

\section{Relative entropy as contrast function}

Let $D_\theta$ be a smooth manifold of density matrices. The following
construction is motivated by classical statistics. Suppose that a 
nonnegative functional $d(D_1,D_2)$ of two variables is given on
the density matrices. In many cases one can get a Riemannian metric
by differentiation:
$$
g_{ij}(\theta)=\frac{\partial^2}{\partial \theta_i \partial \theta_j'}
d(D_\theta, D_{\theta'})\Big|_{\theta=\theta'}
$$
To be more precise the nonnegative smooth functional $d(\pont,\pont)$
is called a contrast functional if $d(D_1,D_2)=0$ iplies $D_1=D_2$.
(For the role of contrast functionals in classical estimation, see 
\cite{Eg}.) We note that a contrast functional is a particular example 
of yokes, cf. \cite{BJ}.

Following the work of Csisz\'ar in classical information theory,
Petz introduced a family of information quantities parametrized by a
function $F:\bbbr^+ \to \bbbr$
\begin{equation}\label{E:Fent}
S_F(D_1,D_2)=\tr(D_1^{1/2}F(\Delta_{D_2,D_1})D_1^{1/2}),
\end{equation}
see \cite{Pz1}, or \cite{O-P} p. 113. Here $\Delta_{D_2,D_1}:=
L_{D_2} R_{D_1}^{-1}$ is the relative modular operator of the two
densities. When $F$ is operator convex, this quasi-entropy possesses
good properties, for example it is a contrast functional in the above
sense if $F$ is not linear. In particular for 
$$
F(t)={4 \over 1-\aa^2}\big(1-t^{(1+\alpha)/2}\big)
$$ 
we have
\begin{equation}\label{E:aaent}
S_\aa (D_1,D_2 )= {4 \over 1-\aa^2}
\tr (I-D_2^{{1+\aa \over 2}}D_1^{-{1+\aa \over 2}})D_1 
\end{equation}
By differentiating we get
\begin{equation}
{\partial^2 \over \partial t \partial u}
S_\aa(D+tA,D+uB)\Big\vert_{t=u=0}=K_D^{\aa}(A,B) 
\end{equation}
which is related to (\ref{E:WYD}) as 
$$
F^\beta_D(A)=K_D^{\aa}(A,A) \qquad \hbox{and}\qquad \beta=(1-\alpha)/2.
$$

Ruskai and Lesniewski discovered that all monotone Fisher informations
are obtained from a quasi-entropy as contrast functional \cite{L-R}.
The relation of the function $F$ in (\ref{E:Fent}) to the function 
$f$ in Theorem \ref{T:mon} is 
\begin{equation}\label{E:Ruskfor}
\frac{1}{f(t)}=\frac{F(t)+tF(t^{-1})}{(t-1)^2}.
\end{equation}

\end{document}